\documentclass[twocolumn,prd,english,american,tightenlines,nofootinbib]{revtex4}
\usepackage[T1]{fontenc}
\usepackage[latin1]{inputenc}
\makeatletter

\def\be{\begin{equation}}
\def\ee{\end{equation}}
\def\beq{\begin{equation}}
\def\eeq{\end{equation}}
\def\bea{\begin{eqnarray}}
\def\eea{\end{eqnarray}}
\def\bml{\begin{subequations}}
\def\blea{\bml\begin{eqnarray}}
\def\elea{\end{eqnarray}\end{subequations}}

\usepackage{babel}
\makeatother
\begin{document}

\title{Towards a non-anthropic solution to the cosmological constant problem}

\author{Andrei Linde}

\email{alinde@stanford.edu}

\author{Vitaly Vanchurin}

\email{vanchurin@stanford.edu}

\date{\today}

\affiliation{Department of Physics, Stanford University, Stanford, CA 94305, USA}

\begin{abstract}

Many probability measures in the multiverse depend exponentially on some observable parameters, giving rise to potential problems such as youngness bias, Q-catastrophe etc. In this paper we explore a possibility that the exponential runaway dependence should be viewed not as a problem, but as a feature that may help us to fix all parameters in the landscape, including the value of the cosmological constant, without using anthropic considerations.
\end{abstract}
\maketitle

\section{Introduction}

Finding the distribution of observables in the multiverse is one of the most important unresolved problems in modern cosmology, see e.g. \cite{Linde:1993nz,Linde:1993xx,GarciaBellido:1993wn,Vilenkin:1994ua,Vanchurin:1999iv,Garriga:2005av,Vanchurin:2006xp,Bousso:2006ev,Linde:2007nm,Winitzki:2008yb,Garriga:2008ks,Linde:2008xf,Noorbala:2010zy}. Continuous work in this area is stimulated by the possibility to merge the theory of eternal inflation and string theory in the context of the string landscape scenario \cite{Vilenkin:1983xq,Linde:1986fd,Lust, Bousso:2000xa, Kachru:2003aw, Douglas:2003um,Susskind:2003kw}. The new cosmological paradigm may help us to solve the cosmological constant problem. More than twenty years ago Weinberg \cite{Weinberg:1987dv} (see also \cite{Linde:1984ir,Sakharov:1984ir,Banks:1984cw,Linde:1986dq}) used the anthropic principle to predict the smallness of the cosmological constant with very mild assumptions about the underlying probability measure. So far this is the only known solution to the cosmological constant problem which we shall refer to as the anthropic solution. In this article we will look for a non-anthropic solution to the problem. 

To proceed we will make two assumptions:

1) The fundamental theory (e.g. string theory) possess only a finite number $N$ of vacuum solutions, landscape of vacua,  with probability distribution  of observable parameters among vacua described by a normalizable function $P(\Lambda,m_{\rm inflaton},...)$.

2) The correct cosmological measure can be described by a positive definite weighting function $w(\Lambda, m_{\rm inflaton},...)$ with an exponential dependence in at least one but possibly many observable parameters. 

The second assumption is quite generic for many probability measures which are exponentially sensitive to the choice of parameters \cite{Garriga:2005ee}. Such measures are often discarded as potentially pathological. In this paper we will take an opposite view, assume that the ``multiverse pressure'' on the choice of the parameters is exponentially strong, and study consequences of this assumption. 

 \section{Runaway observables}\label{runaway}
 
Let vector ${\bf x}=(x,y,z ...)$ denote a random observation in the landscape with $N$ vacua.  Each component of the vector might represent either separate observables (e.g. $x = \Lambda$) or more generally combinations of different observables (e.g. $x = m^{3/2} \Lambda^{-3/4}$). The main idea of our paper is very simple: If we consider a probability measure $w({\bf x})$ that leads to exponentially strong runaway regimes (e.g. Q-catastrophe \cite{Garriga:2005ee}), then the ``multiverse pressure'' related to $w({\bf x})$ may be so powerful that it will unambiguously single out one particular vacuum out of the incredibly large number of other vacua in the landscape. This will simultaneously fix all physical parameters, including the cosmological constant, masses, coupling constants, etc.

We could find the properties of the best vacuum if we would know properties of each vacuum in the landscape, and the corresponding values of $w({\bf x})$:
\be
\langle {\bf x} \rangle = \frac{\sum_{i=1... N} {\bf x}_{i} w({\bf x}_i)}{\sum_{i=1... N} w({\bf x}_i)}.
\label{eq:expectX}
\ee
However, for very large $N$ one might at best hope to derive the probability distribution $P({\bf x})$ of the parameters among different vacua. Thus our main task as cosmologists is to estimate $\langle {\bf x} \rangle$  from the two functions $P({\bf x})$ and  $w({\bf x})$ and an integer $N$. To do so, we must first generate a "random landscape" of $N$ vacua according to $P({\bf x})$ distribution and then populate it with observers according to $w({\bf x})$. Using Eq. (\ref{eq:expectX}) we obtain
\be
\langle {\bf x} \rangle = \int d{\bf x}_1 P({\bf x}_1)... \int  d{\bf x}_N P({\bf x}_N)  \frac{\sum_{i=1... N} {\bf x}_i w({\bf x}_i) }{\sum_{i=1... N} w({\bf x}_i) }.
\label{eq:expectX1}
\ee
Alternatively one might argue that in the limit when observables can vary continuously 
\be
\langle {\bf x} \rangle = \frac{\int d{\bf x} w({\bf x}) P({\bf x}) {\bf x}}{ \int d{\bf x} w({\bf x}) P({\bf x}) }.											\label{eq:expectX2}
\ee
Although Eqs. (\ref{eq:expectX1}) and (\ref{eq:expectX2}) could give similar results, the predictions would differ for $w({\bf x})$ with an exponential pressure in at least one variable.

As was already stated in the introduction, we assume that  the measure-dependent function $w({\bf x})$ is exponentially divergent at some isolated points  ${\bf x}=\tilde{{\bf x}}$ where $P(\tilde{{\bf x}})>0$. Without loss of generality we can always shift the divergent point to the origin $\tilde{{\bf x}} =  (0,0,...)$ and choose the observable parameters such that the weighting function is factorizable near the origin:
\be
w({\bf x} = (x,y,z ...) ) \propto e^{ \frac{C_1}{x}+ \frac{C_2}{y} + \frac{C_3}{z} + ... }
\label{eq:weighting1}
\ee
 where $C_1 \geq C_2 \geq C_3 ...$  and all of the sub-exponential terms were suppressed. Since the values of the coefficients would generically differ  (i.e. $C_1 \neq C_2 \neq C_3 ...$) the Eq. (\ref{eq:expectX1}) can be greatly simplified.  In the limit of infinite $N$ one and only one observational parameter $x$ is obtained from the runaway behavior of the weighting function $\langle x \rangle = 0$  and all other parameters are uniquely determined by only the probability distribution function $P({\bf x})$. For example,
\be
\langle y \rangle = \frac{\int dy dz ... P(\langle x \rangle=0,y,z,...)\,  y}{\int dy dz ... P(\langle x \rangle=0,y,z,...)}.
\label{eq:expectX3}
\ee
What if $N$ is a large, but finite number $\sim10^{500}$ or $10^{120}$? Clearly for a sufficiently large $N$  the  Eq. (\ref{eq:expectX3}) would still hold approximately with one important modification. The expected value of  $x$ would not be exactly zero, but would be slightly shifted. The amount of the shift depends on the probability distribution $P(x)$ of the runaway component and on the number of vacua $N$, but not on the exact shape of $w({\bf x})$.

Now we are ready to tackle the cosmological constant problem. The above analysis suggests that there has to be one and only one observable parameter whose value differs by  many orders of magnitude from all other observables. Among all of the fundamental constants measured so far, the cosmological constant is perhaps the best candidate to play the role of a runaway observable. Its value in dimensionless units is smaller than any other fundamental constant by more than 100 orders in magnitude. This suggests that  the weighting function must depend exponentially on the cosmological constant with $x \equiv f(\Lambda)$:
\be
w(\Lambda) \propto e^{\frac{C_1}{f(\Lambda)}} \ ,
\label{eq:weighting1}
\ee
where $f(\Lambda)$ is a faster than logarithmic function of at least $\Lambda$, but possibly few other parameters. As we will see from some examples below, this divergent behavior may be strong enough to single out one particular vacuum state in the landscape. If we can clearly identify this state, we can unambiguously find all other parameters. However, if our knowledge of the landscape is not good enough, we can still estimate other parameters using the probability distribution $P({\bf x})$ but not $w({\bf x})$.

\section{Examples}\label{examples}
\vskip -2pt
\subsection{Hawking distribution}
\vskip -5pt

As a toy model illustrating this possibility, let us consider the landscape described by a single parameter, the cosmological constant $\Lambda$. Let us assume, as suggested by Hawking long time ago \cite{Hawking:1984hk}, that the cosmological constant is positive and the probability distribution of the cosmological constant in the landscape is given by
\begin{equation}\label{1}
w(\Lambda) \sim e^{S(\Lambda)} = \exp\bigl({24\pi^2/\Lambda}\bigr) \ .
\end{equation}
Here $S(\Lambda)\sim {24\pi^2/ \Lambda}$ is the Euclidean action (entropy) of dS space with  the cosmological constant $\Lambda$. 

A possible intuitive justification of this assumption can be formulated as follows. dS space has entropy $S(\Lambda) ={24\pi^2\over \Lambda}$. An interpretation of dS entropy is a rather challenging endeavour. One may try to interpret it by saying that de Sitter space with a cosmological constant $\Lambda$ is in fact a configuration of $e^{S(\Lambda)}$ universes different from each other at the quantum level. 

Now let us make an assumption based on the Laplace's principle of indifference: {\it a priori}, all quantum states of the universe are equally probable. Since the total number of these states is given by (\ref{1}), this equation represents the probability for the universe with a given value of $\Lambda$. 

We will not attempt to give here a more detailed justification of this probability distribution. In fact, we will shortly provide alternative expressions for $w(\Lambda)$. Our main goal here  is not to identify the best probability measure, but to give a set of simple examples of the mechanism of the non-anthropic selection in the landscape.

In a continuous landscape, the distribution (\ref{1}) leads to a prediction $\Lambda = 0$. However, in the landscape with $N$ vacua the situation is quite different.  Let us assume, for simplicity, that the probability distribution $P$ of the different values of the cosmological constant $\Lambda_{i}$ in the landscape is relatively smooth, which means that $\Lambda$ takes $N$ different values separated from each other by approximately $1/N$ in the interval from $0$ to the Planck density~$1$. Then one may expect that the lowest absolute value of $\Lambda$ will be about $1/N$, the next one will be about $2/N$, etc. How much the lowest value will be rewarded as compared to the next one? Assuming for simplicity that $\Lambda_{1} = 1/N$ and $\Lambda_{2} = 2/N$, one finds that 
\begin{equation}\label{1a}
{w({\Lambda_{2}})\sim  w(\Lambda_{1})}\, \exp\bigl(-{12\pi^2N}\bigr) \ .
\end{equation}
The relative probability to live in a vacuum with $\Lambda \gg \Lambda_{1}$ will be suppressed even stronger, by $\exp\bigl(-{24\pi^2N}\bigr)$. Even if the distribution of values of $\Lambda_{i}$ among possible vacua is not uniform, none of these values can be rewarded by more than the total number of vacua $N$, which is negligible as compared to the reward $\sim \exp\bigl({24\pi^2N}\bigr)$ received by the lowest vacuum in the landscape.

This leads to two important consequences. 

1)  If the values of the cosmological constant are uniformly distributed in the landscape, then the smallest $\Lambda$, which dominates the probability distribution, is given by a simple estimate $\Lambda \sim N^{{-1}}$. If this is the case,\footnote{The possibility that $|\Lambda| \sim N^{{-1}}$ was discussed in the past, see e.g. \cite{Bousso:2009ks,Bousso:2009gx}. However, the probability measure leading to this relation predicted that $\Lambda$ must be negative  \cite{Bousso:2009gx}.}
 one may experimentally determine the total number of vacua in the landscape by measuring $\Lambda$:
\be
N \sim \Lambda^{{-1}} \sim 10^{120} \ , 
\ee
which is not unreasonable \cite{Lust, Bousso:2000xa, Kachru:2003aw, Douglas:2003um,Susskind:2003kw}. This value can by greater than $\Lambda^{{-1}}$ if, for example, the constant $\Lambda$ is uniformly distributed in the interior of an $M$-dimensional sphere of radius $1$. In this case $N \sim\Lambda^{-M} \sim 10^{120\, M}$.

2) If the landscape would contain a continuous variety of vacua, fixing the value of $\Lambda$ would not tell us much about any other parameters. Similarly,  finding $\Lambda$ in the models with a power-law ``multiverse pressure'' is just a first step in the chain of anthropic considerations \cite{Bousso:2009ks,Bousso:2009gx}. However, in the string theory landscape with the exponential dependence of the measure on $\Lambda$, fixing the value of $\Lambda$ singles out one particular vacuum state, rendering all other states irrelevant. This means that by finding the vacuum with the smallest $\left | \Lambda \right |$ we fix all other parameters; no additional anthropic reasoning is required. 

 \subsection{Baby universes reborn}
 \vskip -5pt

Now we will consider another possibility which was very popular in the end of the 80's, and then died, not because it was proven wrong, but because it was extremely difficult to prove it right: the baby universe theory \cite{Coleman:1988cy,Giddings:1988cx,Banks:1988je,Coleman:1988tj}. This theory predicted \cite{Coleman:1988tj} that the probability to live in a universe with a (positive) cosmological constant $\Lambda$ is given by a double exponential expression 
\begin{equation}\label{2}
w(\Lambda) \sim e^{e^{S(\Lambda)}} = \exp\Bigl(\exp\bigl({24\pi^2/\Lambda}\bigr)\Bigr) \ .
\end{equation}
If the cosmological constant could take a continuous set of values, this distribution would imply that $\Lambda = 0$. In the landscape, this distribution would pick up the smallest possible (positive) value of $\Lambda$. Thus, the probability distribution would be much sharper concentrated at the single most probable state than the distribution (\ref{1}), but the predictions in the landscape would be the same as for the Hawking distribution. This would simultaneously fix all other parameters, such as masses and coupling constants. Everything else, including the relation between the minimal value of $\Lambda$ and the total number of vacua in the landscape, also remains the same as in the previous subsection.

\subsection{Counting observable universes}

As we already mentioned, one may try to interpret Eq. (\ref{1}) by saying that de Sitter space with a cosmological constant $\Lambda$ is a configuration consisting of $e^{S(\Lambda)}$ equally probable universes, different from each other at the quantum level. However, one may argue that quantum perturbations in an empty de Sitter space may be important for Boltzmann brains but not for normal classical observers: In a universe which did not experience slow roll inflation, a classical observer will see just one classical vacuum state with a given $\Lambda$ rather than $e^{S(\Lambda)}$ different universes. 

The situation is quite different in a universe which experienced a stage of slow-roll inflation. In such a universe, inflationary quantum fluctuations of metric produce large-scale classical perturbations of geometry which are responsible for CMB anisotropy and large scale structure formation. Therefore for each vacuum state in the landscape, a local observer living in a universe with a given cosmological constant $\Lambda$ may see a large variety of different classical geometries produced during a stage of a slow roll inflation \cite{Linde:2009ah}. What if we consider all of these classical geometries equally probable, and use their number as a tool for evaluation of probabilities?

In ref.  \cite{Linde:2009ah} we found the following estimate for the total number of classical observable universes:
\be
w(N_e) \sim e^{e^{3N_e}} \sim \exp\left(\lambda_{\rm max}/\lambda_{\rm min}\right)^{3}\ .
\label{eq:number}
\ee
Here $N_e$ is the number of e-folds of the slow-roll inflation, $\lambda_{\rm max}/\lambda_{\rm min}$ is  the ratio of the longest wavelength of inflationary perturbations to the smallest wavelength. The origin of this result is related to the fact that the large scale perturbations of the classical geometry are produced during each e-folding, and this process occurs independently during each time interval $O(H^{{-1}})$ in each domain of the size $O(H^{{-1}})$. During inflation $\lambda_{\rm min} \sim H^{{-1}}$, $\lambda_{\rm max} \sim H^{{-1}}e^{N_e}$, but both scales continue growing after inflation. Obviously, this measure exponentially rewards long inflation, which explains why the universe is flat.

However, a local observer can only see a small fraction of these universes. Indeed, if the cosmological constant is negative, the universe collapses within the time $O(|\Lambda|^{-1/2})$, so a local observer can only see a part of the universe on a scale $O(|\Lambda|^{-1/2})$. If the cosmological constant is positive, one cannot see anything beyond the cosmological horizon of the size $|\Lambda|^{-1/2}$. In both cases, $\lambda_{\rm max}$ is limited from above by $|\Lambda|^{-1/2}$. This limit can be reached only if $\lambda_{\rm max} \gtrsim |\Lambda|^{-1/2}$, which explains why in this scenario inflation must be long and the universe must be flat. The scale $|\Lambda|^{-1/2}$ becomes visible only at the time $t \sim |\Lambda|^{-1/2}$, when the density of the ordinary matter decreases and becomes of the same order as $|\Lambda|$, which solves the coincidence problem.  

Let us assume for a moment that $\lambda_{\rm min}$ corresponds to the perturbations produced at the very end of inflation. In this case, an estimate of the maximal number of different locally distinguishable classical geometries made in   \cite{Linde:2009ah} suggests that this number  is given by
\begin{equation}\label{3}
w \sim  \exp\left(H_{I}^{\frac{3}{2}} \left|\Lambda\right|^{-\frac{3}{4}}\right) \ ,
\end{equation}
where $H_{I}$ is the Hubble constant at the end of the slow-roll inflation.

Just as before, the number of the distinguishable geometries grows exponentially when the absolute value of the cosmological constant decreases. In addition, this number also grows for large $H_{I}$, but in string theory the Hubble constant $H_{I}$ cannot be too large, because it destabilizes the vacuum. For example, the value of the inflaton potential in the KKLT scenario typically cannot exceed the depth of the AdS minimum prior to its uplifting, and this depth must be much smaller than $O(1)$  because the AdS minimum appears due to exponentially suppressed non-perturbative effects  \cite{Kallosh:2004yh}.

This may suggest that the inflationary Hubble constant $H_{I}$ must be high, but not too high, which is consistent with the present observational bound  $H_{I} \lesssim 3\times 10^{{-5}}$. This would bring us one step closer to explaining the existence of an incredibly large hierarchy of scales between the two stages of inflation, the one in the very early universe, and the exponential expansion of the universe now, with the Hubble constant $H_{\Lambda} \sim \Lambda^{{1/2}} \sim 10^{{-60}}$. We should note, however, that our conclusions concerning $H_{I}$ are based on the assumption that the $\lambda_{\rm min}$ corresponds to the perturbations produced at the very end of inflation. This is not necessarily the case because the small-scale structure of the universe may be erased by post-inflationary dynamics.  Therefore we would not put too much faith in the $H_{I}$-dependence in Eq. (\ref{3}). Meanwhile the functional dependence on $\Lambda$, which puts the exponential downsizing pressure on $\Lambda$, is much more reliable because it is related to physics on the scale of the cosmological horizon where the nonlinear post-inflationary effects are inessential. Note that this measure does not say anything about the sign of $\Lambda$: The vacuum with the smallest $|\Lambda|$ should win. 

The example discussed above (see also \cite{Bousso:2009ks}) teaches us an interesting lesson. For a general choice of the measure in the exponential runaway scenario, the  most probable vacuum state will be determined by the requirement that a certain function of many different parameters should take its maximal value in the landscape. Maximization of the probability measure (\ref{3}) required the existence of an exponential hierarchy of scales separating $H_{I}$ and $\Lambda^{{1/2}}$. The magnitude of this hierarchy of scales is closely related to the value of $\sqrt N$, where $ N$ is the total number of the vacua in the landscape. Similarly, one may expect that in a more general case the exponential runaway regime discussed in this paper may result in hierarchical relations between other physical parameters.

 \section{The runaway behavior in the landscape and anthropic considerations}\label{anthropic}

The laws of physics in our part of the universe must be compatible with our own existence, hence the anthropic principle. However, the usefulness of this principle depends on the presumed uniqueness of human life and on the relative strength of the anthropic arguments as compared to the runaway behavior of the probability distribution described above.  

The standard approach to the cosmological constant problem is based on the assumption that the prior probability for the cosmological constant is uniform, and the main constraint on the cosmological constant appears because large $\Lambda$ is incompatible with human life. In this paper we are making an opposite assumption. We assume that the prior probability distribution for $\Lambda$ is singular, and the singularity is avoided only because $\Lambda$ can take a discrete set of values. We  also assume that the information processing associated with life is not that rare in the multiverse.

Indeed, we know that our own universe contains many different types of ``observers.'' What if it were inhospitable to humans but quite compatible with observers of some other type? Recent literature on the measure problem in the multiverse contains discussions of a conjecture that our consciousness may be associated with programs which may run on non-biological computers of various nature, see e.g.  \cite{DeSimone:2008if}. It may be possible to significantly modify the laws of physics (i.e. get rid of weak interactions) and yet have a universe which is similar to ours in many important respects, see e.g. \cite{Harnik:2006vj}. In such a universe, ``we'' would not be made of carbon, oxygen and hydrogen, but our properties would be quite compatible with the properties of our universe, which would explain all miraculous coincidences that  we are trying to explain by anthropic considerations.  

 As we have shown in this paper, there are some candidate probability measures in the landscape that single out one particular vacuum and render the probability of all other vacua incredibly small, suppressed by the numbers of the type of $e^{-N}$, where $N$ can be as large as the total number of vacua in the landscape. If such a universe is totally incompatible with any kind of information processing resembling ours, we will turn to the second best universe in the list. In this analysis, the main attention is paid not to the choice of observers, but to the choice of the universe (or the vacuum state in the landscape), in a hope that the probability of emergence of intelligent life is not as small as $10^{-10^{100}}$. Being optimists, we hope that this is a reasonable assumption.

An investigation of the probability measure in the multiverse is complicated and controversial, so we would avoid making any bets here. However, if the scenario outlined in our paper is valid, it will give us a chance to return to the Einstein's dream of a final theory, which may allow us to make sharp and unambiguous predictions despite the abundance of choice.\\
\vskip 15pt

{\it Acknowledgments.} The authors are grateful to R. Kallosh, V. Mukhanov, M. Noorbala and M. Salem  for helpful discussions. The work of A. L. was supported in part by NSF grant PHY-0756174 and  by the FQXi grant RFP2-08-19. The work of V. V. was supported in part by FQXi mini-grants MGB-07-018 and MGA-09-017.


\begin{thebibliography}{10}

\bibitem{Linde:1993nz}
  A.~D.~Linde and A.~Mezhlumian, ``Stationary universe,''
  Phys.\ Lett.\  B {\bf 307}, 25 (1993)  
  [arXiv:gr-qc/9304015].
  
\bibitem{Linde:1993xx}
  A.~D.~Linde, D.~A.~Linde and A.~Mezhlumian,
  ``From the Big Bang theory to the theory of a stationary universe,''
  Phys.\ Rev.\  {\bf D49}, 1783-1826 (1994).
  [gr-qc/9306035].
  
\bibitem{GarciaBellido:1993wn}
  J.~Garcia-Bellido, A.~D.~Linde and D.~A.~Linde,
  ``Fluctuations of the gravitational constant in the inflationary Brans-Dicke cosmology,''
  Phys.\ Rev.\  {\bf D50}, 730-750 (1994).
  [astro-ph/9312039]. 
  
  \bibitem{Vilenkin:1994ua}
  A.~Vilenkin,
``Predictions from quantum cosmology,''
  Phys.\ Rev.\ Lett.\  {\bf 74}, 846 (1995)
  [arXiv:gr-qc/9406010].

\bibitem{Vanchurin:1999iv}
  V.~Vanchurin, A.~Vilenkin and S.~Winitzki,
``Predictability crisis in inflationary cosmology and its resolution,''
  Phys.\ Rev.\  {\bf D61}, 083507 (2000).
  [gr-qc/9905097].

\bibitem{Garriga:2005av}
  J.~Garriga, D.~Schwartz-Perlov, A.~Vilenkin and S.~Winitzki,
``Probabilities in the inflationary multiverse,''
  JCAP {\bf 0601}, 017 (2006)
  [arXiv:hep-th/0509184].
  
\bibitem{Vanchurin:2006xp}
  V.~Vanchurin,
``Geodesic measures of the landscape,''
  Phys.\ Rev.\  {\bf D75}, 023524 (2007).
  [hep-th/0612215]. 
  
 \bibitem{Bousso:2006ev}
  R.~Bousso,
``Holographic probabilities in eternal inflation,''
  Phys.\ Rev.\ Lett.\  {\bf 97}, 191302 (2006).
  [hep-th/0605263]. 
  
  \bibitem{Linde:2007nm}
  A.~D.~Linde,
``Towards a gauge invariant volume-weighted probability measure for eternal inflation,''
  JCAP {\bf 0706}, 017 (2007).
  [arXiv:0705.1160 [hep-th]].  
  
  \bibitem{Winitzki:2008yb}
  S.~Winitzki,
``A Volume-weighted measure for eternal inflation,''
  Phys.\ Rev.\  {\bf D78}, 043501 (2008).
  [arXiv:0803.1300 [gr-qc]]. 
  
\bibitem{Garriga:2008ks}
  J.~Garriga and A.~Vilenkin,
``Holographic Multiverse,''
  JCAP {\bf 0901}, 021 (2009).
  [arXiv:0809.4257 [hep-th]]. 
  
  \bibitem{Linde:2008xf}
  A.~D.~Linde, V.~Vanchurin, S.~Winitzki,
``Stationary Measure in the Multiverse,''
  JCAP {\bf 0901}, 031 (2009).
  [arXiv:0812.0005 [hep-th]].  
  
  \bibitem{Noorbala:2010zy}
  M.~Noorbala and V.~Vanchurin,
``Geocentric cosmology: A New look at the measure problem,''
  [arXiv:1006.4148 [hep-th]].
  
\bibitem{Vilenkin:1983xq}
  A.~Vilenkin,
``The Birth Of Inflationary Universes,''
  Phys.\ Rev.\  D {\bf 27}, 2848 (1983).

\bibitem{Linde:1986fd}
  A.~D.~Linde,
``Eternally Existing Self-reproducing Chaotic Inflationary Universe,''
  Phys.\ Lett.\  B {\bf 175}, 395 (1986).
  
   \bibitem{Lust}  
 W.~Lerche, D.~L\"ust and A.~N.~Schellekens,
``Chiral Four-Dimensional Heterotic Strings From Selfdual Lattices,''
  Nucl.\ Phys.\ B {\bf 287}, 477 (1987).

\bibitem{Bousso:2000xa}
  R.~Bousso and J.~Polchinski,
 ``Quantization of four form fluxes and dynamical neutralization of the cosmological constant,''
  JHEP {\bf 0006}, 006 (2000).
  [hep-th/0004134].
  
  
\bibitem{Kachru:2003aw}
S.~Kachru, R.~Kallosh, A.~Linde and S.~P.~Trivedi, 
``De Sitter vacua in string theory,'' 
Phys.\ Rev.\ D {\bf 68}, 046005 (2003).
[arXiv:hep-th/0301240].

\bibitem{Douglas:2003um}
  M.~R.~Douglas,
  ``The Statistics of string / M theory vacua,''
  JHEP {\bf 0305}, 046 (2003).
  [hep-th/0303194].
  
 \bibitem{Susskind:2003kw}
  L.~Susskind,
  ``The anthropic landscape of string theory,''
  arXiv:hep-th/0302219.

  
\bibitem{Weinberg:1987dv}
  S.~Weinberg,
``Anthropic Bound on the Cosmological Constant,''
  Phys.\ Rev.\ Lett.\  {\bf 59}, 2607 (1987).

  
\bibitem{Linde:1984ir}
A.~D.~Linde,
``The Inflationary Universe,''
  Rept.\ Prog.\ Phys.\  {\bf 47}, 925 (1984).  
  
  \bibitem{Sakharov:1984ir}
  A.~D.~Sakharov,
  ``Cosmological Transitions With A Change In Metric Signature,''
  Sov.\ Phys.\ JETP {\bf 60}, 214-218 (1984).
  [Zh.\ Eksp.\ Teor.\ Fiz.\  {\bf 87}, 375 (1984)].  
  
\bibitem{Banks:1984cw}
  T.~Banks,
  ``T C P, Quantum Gravity, the Cosmological Constant and All That...,''
  Nucl.\ Phys.\  {\bf B249}, 332 (1985).
  
\bibitem{Linde:1986dq}
  A.~D.~Linde, 
``Inflation And Quantum Cosmology,'' 
in: {\it  Three hundred years of gravitation}.  Cambridge Univ. Press, Eds.  
Hawking, S.W.  and Israel, W., 604-630 (1987).




   \bibitem{Garriga:2005ee}
  J.~Garriga and A.~Vilenkin,
``Anthropic prediction for Lambda and the Q catastrophe,''
  Prog.\ Theor.\ Phys.\ Suppl.\  {\bf 163}, 245 (2006).
 [hep-th/0508005].
 
  
\bibitem{Hawking:1984hk}
  S.~W.~Hawking,
``The Cosmological Constant Is Probably Zero,''
  Phys.\ Lett.\  B {\bf 134}, 403 (1984).
  

\bibitem{Bousso:2009ks}
  R.~Bousso, L.~J.~Hall and Y.~Nomura,
``Multiverse Understanding of Cosmological Coincidences,''
  Phys.\ Rev.\  D {\bf 80}, 063510 (2009)
  [arXiv:0902.2263 [hep-th]].
 
 
\bibitem{Bousso:2009gx}
  R.~Bousso and S.~Leichenauer,
  ``Predictions from Star Formation in the Multiverse,''
  Phys.\ Rev.\  D {\bf 81}, 063524 (2010).
  [arXiv:0907.4917 [hep-th]].
 

 

  
 
\bibitem{Coleman:1988cy}
  S.~R.~Coleman,
 ``Black Holes as Red Herrings: Topological Fluctuations and the Loss of
Quantum Coherence,''
  Nucl.\ Phys.\  B {\bf 307}, 867 (1988).  
  
 \bibitem{Giddings:1988cx}
  S.~B.~Giddings and A.~Strominger,
``Loss of Incoherence and Determination of Coupling Constants in Quantum
Gravity,''
  Nucl.\ Phys.\  B {\bf 307}, 854 (1988).


\bibitem{Banks:1988je}
  T.~Banks,
  ``Prolegomena to a Theory of Bifurcating Universes: A Nonlocal Solution to the Cosmological Constant Problem Or Little Lambda Goes Back to the Future,''
  Nucl.\ Phys.\  {\bf B309}, 493 (1988).
  
 \bibitem{Coleman:1988tj}
  S.~R.~Coleman,
 ``Why There Is Nothing Rather Than Something: A Theory of the Cosmological Constant,''
  Nucl.\ Phys.\  B {\bf 310}, 643 (1988).
  
\bibitem{Linde:2009ah}
 A.~Linde and V.~Vanchurin,
 ``How many universes are in the multiverse?,''
  Phys.\ Rev.\  {\bf D81}, 083525 (2010).
  [arXiv:0910.1589 [hep-th]].
  

\bibitem{Kallosh:2004yh}
  R.~Kallosh and A.~D.~Linde,
 ``Landscape, the scale of SUSY breaking, and inflation,''
  JHEP {\bf 0412}, 004 (2004)
  [arXiv:hep-th/0411011].


\bibitem{DeSimone:2008if}
  A.~De Simone, A.~H.~Guth, A.~D.~Linde, M.~Noorbala, M.~P.~Salem and A.~Vilenkin,
``Boltzmann brains and the scale-factor cutoff measure of the multiverse,''
  Phys.\ Rev.\  D {\bf 82}, 063520 (2010).
  [arXiv:0808.3778 [hep-th]].
  
\bibitem{Harnik:2006vj}
  R.~Harnik, G.~D.~Kribs and G.~Perez,
``A universe without weak interactions,''
  Phys.\ Rev.\  D {\bf 74}, 035006 (2006).
  [arXiv:hep-ph/0604027].

\end{thebibliography}
\end{document}